\documentclass[3p,times]{article}

\usepackage{amssymb}
\usepackage{amsthm}
\usepackage{amsmath}
\usepackage{authblk}

\usepackage[figuresright]{rotating}

\def\d{\mathrm{d}}

\begin{document}

\title{$B\to\rho$, $K^*$ transition form factors in AdS/QCD model.}

\author[1]{Mohammad Ahmady\thanks{mahmady@mta.ca (Presenting author)}}
\author[1]{Robyn Campbell\footnote{Now at: Department of Physics, Carleton University, Ottawa, Ontario, Canada, K1S 5B6}\thanks{recampbell@mta.ca}}
\author[2]{S\'{e}bastien Lord\thanks{esl8420@umoncton.ca}}
\author[1,3]{Ruben Sandapen\thanks{ruben.sandapen@umoncton.ca}}
\affil[1]{\small Department of Physics, Mount Allison University, \mbox{Sackville, New Brunswick, Canada, E4L 1E6}}
\affil[2]{D\'epartement de Math\'ematiques et Statistique, Universit\'{e} de Moncton, \mbox{Moncton, New Brunswick, Canada, E1A 3E9}}
\affil[3]{D\'epartement de Physique et d'Astronomie, Universit\'{e} de Moncton, \mbox{Moncton, New Brunswick, Canada, E1A 3E9}}
\renewcommand\Authands{ and }

\maketitle

\begin{abstract}
We use the AdS/QCD distribution amplitudes for light mesons to calculate the transition form factors for B to $\rho$, $K^*$ decays.  These form factors are then utilized to make predictions for the semileptonic $B\to\rho\ell\nu$ and dileptonic $B\to K^* \mu^+\mu^-$ decays.  We compare our predictions to the experimental data from BaBar and LHCb.
\end{abstract}

\section{Introduction}
\label{}
Semileptonic $B$ decays like $B \to \rho \ell \nu$ and $B \to K^* \mu^+ \mu^-$ are very useful for precision tests of the Standard Model and for probing New Physics (NP). To fully understand such decays, it is essential  to model the strong interaction between the quark and antiquark forming the final light meson.  This is a challenging non-perturbative problem which we address here using a relatively new tool named the anti-de Sitter/Quantum Chromodynamics (AdS/QCD) correspondence.
A remarkable feature of the AdS/QCD correspondence, recently discovered by Brodsky and de T\'eramond, is referred to as light-front holography \cite{Brodsky,physicsreport}. In light-front QCD, with massless quarks, the meson wavefunction can be written in the following factorized form \cite{Brodsky}: 
\begin{equation}
	\phi(z,\zeta, \varphi)=\frac{\Phi(\zeta)}{\sqrt{2\pi \zeta}} f(z) \mathrm{e}^{i L \varphi}
	\label{factorized-lc}
\end{equation}
with $\Phi(\zeta)$ satisfying the so-called holographic light-front Schroedinger equation (hLFSE)
\begin{equation}
	\left(-\frac{d^{2}}{d\zeta^2}-\frac{1-4L^{2}}{4\zeta^{2}}+U(\zeta)\right)\Phi(\zeta)=M^{2}\Phi(\zeta) \;.
\end{equation}
In the above equations, $L$ is the orbital angular momentum quantum number and $M$ is the mass of the meson. The variable $\zeta=\sqrt{z(1-z)}r$ where $r$ is the transverse distance between the quark and antiquark forming the meson and $z$ is the fraction of the meson's momentum carried by the quark. Remarkably, the hLFSE maps onto the wave equation for strings propagating in AdS space if $\zeta$ is identified with the fifth dimension in AdS.  The confining potential $U(\zeta)$ in physical spacetime is then determined by the perturbed geometry of AdS space.  In particular, a quadratic dilaton field breaking the conformal invariance of AdS space yields a harmonic oscillator potential in ordinary spacetime, i.e.
\begin{equation}
	U(\zeta)=\kappa^4 \zeta^2 + 2\kappa^2(J-1)
	\label{quadratic-dilaton}
\end{equation}
where $J=L+S$. The holographic light-front wavefunction for a vector meson $(L=0,S=1)$ then becomes
{
	\begin{equation}
		\phi_{\lambda} (z,\zeta) \propto \hspace{-0.1cm} \sqrt{z(1-z)} \exp
		\left(-\frac{\kappa^2 \zeta^2}{2}\right)
		\exp\left \{-\left[\frac{m_q^2-z(m_q^2-m^2_{\bar{q}})}{2\kappa^2 z (1-z)} \right]
		\right \} \label{AdS-QCD-wfn}
	\end{equation}
}
with $\kappa=M_{V}/\sqrt{2}$ and where the dependence on quark masses has been introduced using a prescription by Brodsky and de T\'eramond. This holographic wavefunction has been used to successfully  predict diffractive $\rho$-meson electroproduction \cite{PRL}. 
\section{Distribution amplitudes}
\begin{figure}
\begin{center}
	\includegraphics[width=0.8\linewidth, trim = 0cm 0cm 0cm 1cm, clip]{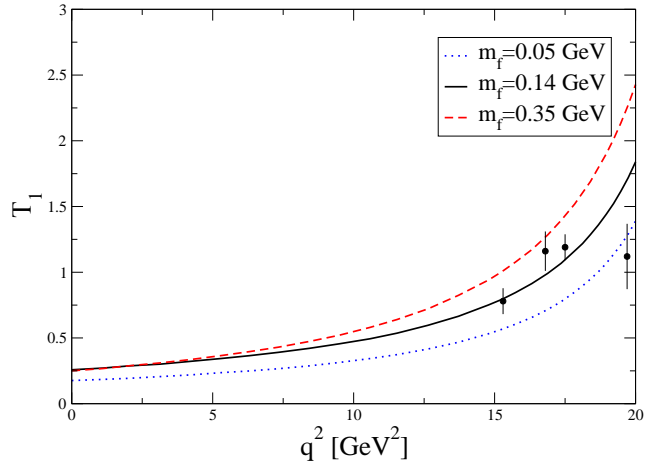}
	\caption{Our predictions for the form factor $T^{B\to \rho}_1$\cite{PRD3} compared to the lattice data \cite{UKQCD}.}
	\label{T1rho}
\end{center}
\end{figure}

\begin{figure}
\begin{center}
	\includegraphics[width=0.8\linewidth, trim = 0cm 0cm 0cm 1cm, clip]{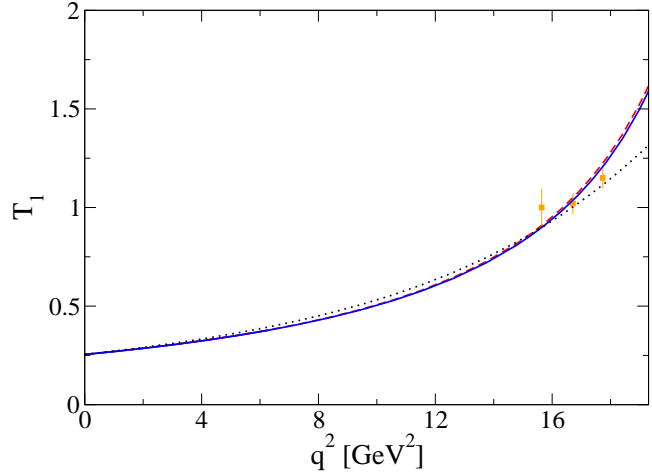}
	\caption{Our predictions for the form factor $T^{B\to K*}_1$\cite{PRD4} and the lattice data \cite{latticekstar}.}
	\label{T1Kstar}
\end{center}
\end{figure}

The Distribution Amplitudes (DAs) of the meson are related to its light-front wavefunction which in turn can be obtained using AdS/QCD.  For vector mesons such as $\rho$ or $K^{*}$, there are two such DAs at twist-$2$ accuracy. The DAs are important because they are inputs in the light-cone sum rules computations of $B \to \rho$ and $B \to  K^{*}$ transition form factors. These form factors are in turn required to compute the decay rates of $B \to \rho \ell \nu$ and $B \to K^{*} \mu^{+} \mu^{-}$.  The leading twist-$2$  DAs for vector mesons are related to the meson light-front wavefunction as follows \cite{PRD1,PRD2}:

	{
		\begin{equation}
		\! \! \! \! \phi_{V}^\parallel(z,\mu) \propto \int \d
		r \; \mu
		J_1(\mu r) [M_{V}^2 z(1-z) + m_qm_{\bar q} -\nabla_r^2] \frac{\phi_L(r,z)}{z(1-z)} \;
		\label{phiparallel-phiL}
		\end{equation}
		\begin{equation}
		\phi_{V}^{\perp}(z,\mu) \propto \int \d
		r \; \mu
		J_1(\mu r)[(1-z)m_q+zm_{\bar q}] \frac{\phi_T(r,z)}{z(1-z)} \;
		\label{phiperp-phiT}
		\end{equation}
	}
The form factors are then obtained from the DAs using the light cone sum rules as explained in reference \cite{ali}. For example, our predictions for the tensor form factors $T^{B \to V}_1$, as a function of the momentum transfer squared, are shown in figures \ref{T1rho} and \ref{T1Kstar}.  

\section{Results}
The total differential decay width for  $B\to\rho\ell\nu$ is given by
\begin{equation}
\frac{\d \Gamma}{\d q^2} =
\frac{G_F^2|V_{ub}|^2}{192\pi^3m_B^3}\,\sqrt{\lambda(q^2)} \,q^2\left(H_0^2(q^2) +
H_+^2(q^2)+ H_-^2(q^2)\right)
\end{equation}
where
\begin{equation}
\lambda(q^2)=(m_B^2 + m_{\rho}^2 -q^2)^2 -4 m_B^2 m_\rho^2 
\end{equation}
and transverse and longitudinal helicity amplitudes for the decay $B \to \rho l \nu$ are given by \cite{Ball:1997rj}
\begin{equation}
H_{\pm}(q^2)=(m_B + m_{\rho}) A_1(q^2) \mp \frac{\sqrt{\lambda(q^2)}}{m_B + m_\rho} V(q^2)
\end{equation}
and
\begin{equation}
H_0 (q^2) = \frac{1}{2m_\rho\sqrt{q^2}}\bigg\{(m_B^2 - m_\rho^2 - t) (m_B + m_\rho) A_1(q^2) - \frac{\lambda (q^2)}{m_B + m_\rho}A_2(q^2)\bigg\}
\end{equation}

respectively. $V$, $A_1$ and $A_2$ are $B\to \rho$ transition form factors which we calculate using AdS/QCD DAs.

The BaBar collaboration has measured partial decay widths in three different $q^2$ bins\cite{babar}:

\begin{center}
	{\begin{tabular}{| c | c | c |}
			
			\hline
			
			{\small $\Delta \Gamma_{\mbox{\tiny{low}}}= \int_0^8 \frac{d \Gamma}{d q^2} d q^2 $} &
			{\small $\Delta \Gamma_{\mbox{\tiny{mid}}}= \int_8^{16} \frac{d \Gamma}{d q^2} d q^2$} & 
			{\small $\Delta \Gamma_{\mbox{\tiny{high}}}= \int_{16}^{20.3} \frac{d \Gamma}{ q^2}  q^2$} \\
			
			\hline
			
			{\small $(0.564 \! \pm \! 0.166) \! \cdot \! 10^{-4}$} &
			{\small $(0.912 \! \pm \! 0.147) \! \cdot \! 10^{-4}$} &
			{\small $(0.268 \! \pm \! 0.062) \! \cdot \! 10^{-4}$} \\
			
			\hline
			
		\end{tabular}}
	\end{center}
	so that
	{\small
		\begin{eqnarray}
		R_{\mbox{\tiny{low}}}=\frac{\Delta\Gamma_{\mbox{\tiny{low}}}}{\Delta\Gamma_{\mbox{\tiny{mid}}}}=0.618 \pm 0.207 \nonumber
		\hspace{0.5cm}
		R_{\mbox{\tiny{high}}}=\frac{\Delta\Gamma_{\mbox{\tiny{high}}}}{\Delta\Gamma_{\mbox{\tiny{mid}}}}=0.294 \pm 0.083
		\end{eqnarray}
	}
	Our predictions for the above ratios are:
	\begin{center}
		$R_{\mbox{\tiny{low}}}=0.580, 0.424$
		\hspace{1cm}
		$R_{\mbox{\tiny{high}}}=0.427,0.503$
	\end{center}
	respectively, for two different quark mass values $m_q=0.14,\; 0.35$ GeV.

	We shall now consider $B\to K^*\mu^+\mu^-$ dileptonic decay mode.  We first compute $B\to K^*$ form factors for low to intermediate values of $q^2$ using AdS/QCD DAs.  For each form factor, we fit the parametric form 
	\begin{equation}
	F(q^2)=\frac{F(0)}{1- a (q^2/m_B^2) + b (q^4/m_B^4)}
	\label{FitFF}
	\end{equation}
	to our predictions. The fitted values of the parameters $a$ and $b$ are given in Table \ref{tab:abAdS}. 
	We repeat the fits by including  the most recent unquenched lattice data  of Ref. \cite{horgan}. The fitted values of $a$ and $b$ are collected in Table \ref{tab:abAdSlat}.

	\begin{table}[h]
		\begin{center}
			\[
			\begin{array}
			[c]{|c|c|c|c|}\hline
			F&F(0)&a&b\\ \hline
			A_0&0.285& 1.158 & 0.096 \\ \hline
			A_1&0.249& 0.625 &-0.119  \\ \hline
			A_2&0.235&  1.438&0.554 \\ \hline
			V&0.277& 1.642& 0.600\\ \hline
			T_1&0.255& 1.557& 0.499\\ \hline
			T_2 & 0.251 &0.665  &-0.028  \\ \hline
			T_3& 0.155 & 1.503&0.695 \\ \hline
			\end{array}
			\]
		\end{center}
		\caption {The values of the form factors at $q^2=0$ together with the fitted parameters $a$ and $b$. The values of $a$ and $b$ are obtained by fitting to the AdS/QCD predictions for low-to-intermediate $q^2$. }
		\label{tab:abAdS}
	\end{table}
	
	\begin{table}[h]
		\begin{center}
			\[
			\begin{array}
			[c]{|c|c|c|c|}\hline
			F&F(0)&a&b\\ \hline
			A_0&0.285& 1.314&0.160 \\ \hline
			A_1&0.249&  0.537&-0.403 \\ \hline
			A_2&0.235& 1.895&1.453 \\ \hline
			V&0.277& 1.783&0.840\\ \hline
			T_1&0.255&1.750 &0.842 \\ \hline
			T_2 & 0.251 &0.555 &-0.379\\ \hline
			T_3& 0.155 &1.208 &-0.030 \\ \hline
			\end{array}
			\]
		\end{center}
	\caption {The values of the form factors at $q^2=0$ together with the fitted parameters $a$ and $b$. The values of $a$ and $b$ are obtained by fitting to both the AdS/QCD predictions for low-to-intermediate $q^2$ and the lattice data at high $q^2$. }
	\label{tab:abAdSlat}
\end{table}

	We use the formula given in Ref \cite{Aliev} to predict the differential branching ratio.  Our predictions are compared with the LHCb data in figure 3.
	By integrating over $q^2$ and excluding the regions of the narrow charmonium resonances, we obtain a total branching fraction of $1.56 \times 10^{-6}(1.55 \times 10^{-6})$ when we are including (excluding) the lattice data compared to the LHCb measurement $(1.16 \pm 0.19)\times 10^{-6}$. In both cases, we overestimate the total branching fraction. With a new physics contribution to $C_9$, we obtain a total branching fraction of $1.35 \times 10^{-6}$ in agreement with the LHCb data.
	\section{Conclusion}
	Light-front holography is a new remarkable feature of the AdS/QCD correspondence. We have used it here to compute the non-perturbative Distribution Amplitudes for the vector mesons $\rho$ and $K^*$ which we then use as inputs to light-cone sum rules to predict $B \to V$ transition form factors. Agreement with data is good especially for low to moderate momentum transfer.
	
	\begin{figure}
	\begin{center}
		\includegraphics[width=0.9\linewidth]{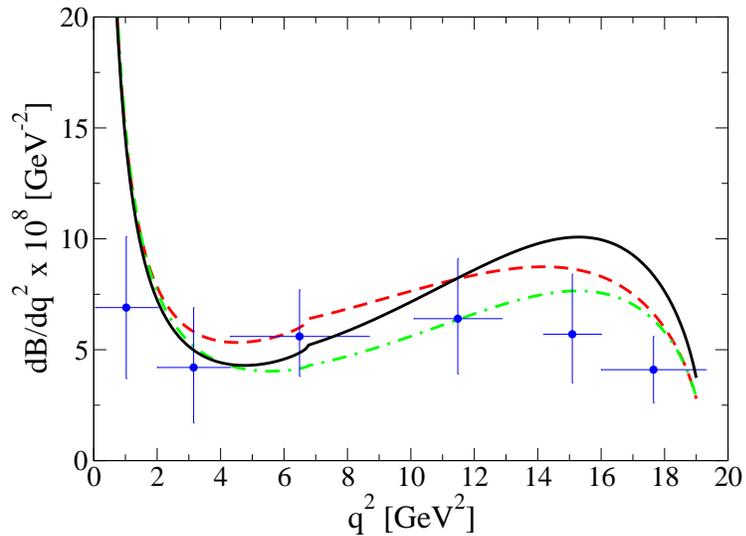}
		\caption{Differential branching ratio: Dashed-red AdS/QCD, solid-black AdS+lattice, green AdS+lattice and new physics.  The data from LHCb\cite{LHCb}.}
\end{center}
	\end{figure}
	
	\section{Acknowledgment}
	This research is supported by a team Discovery Grant from Natural Sciences and Engineering Research Council of Canada (NSERC).  The presentation at ICHEP 2014 is supported in part by a Marjorie Young Bell Award from Mount Allison University.

\end{document}